\documentclass[epj]{svjour}
\usepackage{graphicx}
\usepackage{bm}   
\usepackage{color}
\usepackage{amssymb}
\usepackage{amsmath}
\usepackage{balance}
\usepackage{subfigure}

\begin{document}
	
\title{Penrose Instabilities and the Emergence of Rogue waves in Sasa-Satsuma equation}
\author{M. Pradeepa\inst{1} \and N. Vishnu Priya\inst{2}  \and M. Senthilvelan\inst{1}\mail{velan@cnld.bdu.ac.in}
}                     
%
%
\institute{Department of Nonlinear Dynamics, Bharathidasan University, Tiruchirappalli 620024, Tamil Nadu, India. \and Department of Mathematics, Indian Institute of Science, Bangalore-560012, Karnataka, India.}
\date{Received: -- / Revised version: --}
%
\abstract{
	 In this paper, we calculate the region of emergence of rogue waves in the Sasa-Satsuma equation by performing Penrose stability analysis. We consider Wigner-transformed Sasa-Satsuma equation and separate out unstable solutions, namely Penrose instability modes. We superpose these modes in a small region. With the help of marginal property of the Wigner transform we  identify the region in which rogue wave solution can emerge in the Sasa-Satsuma equation and calculate the amount of spatial localization. We also formulate a condition for the emergence of rogue wave solution in the Sasa-Satsuma equation. 
	\PACS{
		{PACS-key}{Rogue wave \and Penrose instability \and Wigner transform}
	} 
}
\titlerunning{Penrose Instabilities and the Emergence of Rogue waves in Sasa-Satsuma equation}
\authorrunning{M. Pradeepa et al.}
\maketitle
\section{Introduction}
\par Modulation instability (MI), which unties the nonlinear mysteries, has been studied in  depth in various nonlinear systems that arise in different branches of Physics including hydrodynamics \cite{OKim}, plasma physics \cite{YHICH} and optics \cite{Mei}. A very simple mathematical model which has undergone this analysis is the nonlinear Schr\"{o}dinger equation which is represented mathematically by the nonlinear partial differential equation,
\begin{equation}
	iq_t+\frac{1}{2}q_{xx}+\lvert q \rvert^2q=0,
	\label{nls}
\end{equation}
where $q$ is a complex wave envelope, $x$ and $t$ are space and time co-ordinates, subscripts represent partial derivative. The second and third term in Eq. (\ref{nls}) refer to  group velocity dispersion and self-phase modulation respectively. The  characteristics of MI for a family of NLS equations has  been a topic of research over decades \cite{JHZha,MicRem}. The MI  studies have also been carried out on collisionless electron-ion plasma wave in order to explain the  ordinary Landau damping in that system \cite{YHICH}. Earlier Penrose investigated the instability  of the Vlasov-Poisson equation \cite{Pen60}. 
\par The MI can explain the resonance generation which occurs in the nonlinear system that comes from self interacting waves \cite{Rybak}. In the literature, MI is also known as Benjamin and Feir instability \cite{TJ}. Lighthill introduced a geometrical condition for the instability of deep water waves, as a development of Whitham's theory \cite{MJ}.  Later, Zakharov  ventured into Hamiltonian approach and derived the MI criterion for  water waves \cite{VEZ}. MI may develop due to  self induced effect or due to induced one. When a weak external amplitude  or phase perturbation is applied to an initial continuous wave, MI can be induced if the sideband wavenumber falls within the critical wavenumber. The instability of this kind is called  induced MI. If the modulation develops spontaneously due to  noise that is present in the system then it is called spontaneous  MI\cite{MicRem}. 
Regardless whether it is self induced or induced one, MI plays a vital role in the formation of localized structures including solitons of various types, breathers and rogue waves (RWs) \cite{Mono,Akh}. These localized structures are frequently observed both in experiments and in theory. In this work, we focus on the instabilities which lead to the formation of RWs.
\par    In a recent work, Athanassoulis et al. have studied Penrose instabilies in the NLS equation  and built RW from the spatially periodic MI-type modes \cite{Athan}. MI is observed on a plane wave background.  Unlike MI, the Penrose instability can be observed on a spectral background. Since the results of \cite{Athan} are obtained for the narrow spectral conditions, it generalizes the MI. The authors of \cite{Athan} have also calculated the fundamental length scales and timescales for the localization of the above said modes to raise as RWs. In the Penrose stability analysis, the spectral background  can be brought inside the theory through Wigner transform.  Based on the work \cite{Athan}, recently a higher order NLS equation, namely Hirota equation, is analyzed where the contribution of higher order nonlinear terms in the Lorentian spectrum and damping have been investigated \cite{Al-Ta}.  
\par In this work, we consider an extended version of NLS equation, namely Sasa-Satsuma equation (SSE), of the form
\begin{equation}
	iq_t+\frac{q_{xx}}{2}+\lvert q \rvert^2q-i\gamma\Big[q_{xxx}+3\big(\lvert q \rvert^2\big)_x q+6\lvert q\rvert^2q_x\Big]=0,
	\label{e1}
\end{equation}
where $q$, $x$, $t$ and the subscripts mean the same as in Eq. (\ref{nls}), and  $\gamma$ is the strength of the higher order terms. Equation (\ref{e1}) acts as an important model to describe third order dispersion, self frequency shift, and self-steepening effect of the nonlinear dispersive waves. These higher order nonlinearities constitute structural complexities in the solution of SSE. A number of publications have been made to construct the solutions of SSE \cite{NA,UBC,UbanLet,UbanRev}. These exact analytical solutions represent events like breathers, solitons and RWs that occur in various nonlinear media. Besides these, one can also find several other interesting localized structures in the SSE (\ref{e1}) \cite{UbanRev,UbanLet,Chens,Li16,JJC}. In this paper, we study  Penrose instability  and identify the region of the emergence of RW in the SSE. We also aim to quantify the spatial localization of RW. 
\par To achieve these two goals, to begin, we rewrite the SSE in phase space using Wigner transform and its properties. We  investigate the stability of the constructed Wigner-SSE. By feebly perturbing the homogeneous spectrum we obtain a linearize Wigner-SSE. By considering a physically valid solution to the linearized Wigner-SSE we formulate the Penrose instability condition. We then evaluate the obtained Penrose instability condition by  considering the background as a narrow spectrum and separate out the unstable modes that  arise from them. By continuously superposing the unstable modes we obtain a localized structure which we call localized instabilities. We then calculate the exact amount of spatial localization of these instabilities. The occurrence of RW has been pointed out in the literature as ``a linear unstable mode localized over a single wavelength $\lambda_0=2\pi/k_0$". Hence the  localized instabilities which we calculate should be within the region $\lambda_0$, from which we derive a condition for the occurrence of RW. The RW solution reported earlier in the literature satisfies the condition which we  derive in this paper. We note here that in this analysis  any type of Fourier spectrum can be chosen as the background to determine the RW emergence. However, MI can be observed only on the plane wave background, for more details see Ref. \cite{Athan}.

\par We organize our work as follows: In Sec. \ref{S2}, we recall briefly the Wigner transform and its properties and construct the SSE in phase space using the Wigner transform. In Sec. \ref{S3}, we linearize the nonlinear SSE. We perform Penrose instability analysis and formulate the Penrose instability criterion. We determine the unstable solutions of the Wigner-SSE, arising from the homogeneous spectral background in the narrow band limit we identify unstable modes from these we obtain localized instabilities. In Sec. \ref{S4}, we superpose these unstable modes and create a  localized structure in space and  quantify this localization. Using the RW criteria given in \cite{Athan}, we calculate the region where the RW can emerge. In Sec. \ref{S5}, we summarize our work and highlight the advantages and differences of our work with earlier works. In Appendix \ref{AppA}, we derive the Wigner-SSE.

\section{\label{S2} Construction of  Wigner-SSE}
\par In this section, to begin, we recall Wigner transform and  some of its essential properties. We  rewrite the SSE in phase space with the help of Wigner transform and its properties. We call the SSE which is written in phase space as Wigner-SSE. 
\subsection{\label{SS21}Wigner Transform}
\par The Wigner transform was formulated by Eugene Paul Wigner\cite{Wig}.  The motivation to introduce this transform is to make a quantal corrections to classical statistics \cite{Wig}. Now Wigner transfrom is being used in various fields \cite{Wein,Wil08}. In this work, we choose Wigner transform for the dynamical study of SSE. The Wigner transform of a complex wave envelope is defined by \cite{Athan,Athan2}
\begin{equation}
	W(x,k,t)=W[q,q]=\int_{s}e^{-2\pi iks}q\big(x+\frac{s}{2},t\big)~\bar q\big(x-\frac{s}{2},t\big)~ds,
	\label{e2}
\end{equation}
where  $W(x,k,t)$ is the Wigner transform, $q$ is the complex wave envelope,  $k$ is the wavenumber, $x$ and $t$ are space and time coordinates respectively and $bar$ denotes the complex conjugation. Equation (\ref{e2}) is the sub-case of the conventional Wigner transform
\begin{equation}
	W(x,k,t)=W[q,p]=\int_{s}e^{-2\pi iks}q\big(x+\frac{s}{2},t\big)~\bar p\big(x-\frac{s}{2},t\big)~ds,
	\label{e3}
\end{equation}
where $p$ is another complex wave envelope. The integration is over the entire space. As our aim is to construct Wigner transform for the evolutionary  Eq. (\ref{e1}) which contains the variable $q$ and its complex conjugate we consider Wigner transform in the form (\ref{e2}). Since the nonlinear PDE  (\ref{e1}) contains derivative terms as well, in the following, we recall a couple of  properties of Wigner transform which are essential  to reformulate  Eq. (\ref{e1}) in phase space.\\
(i) {\bf Marginals:} When a continuous multivariate function is integrated over one of its variates, then the integral is said to be its marginal distribution if the result is a probability distribution  in the remaining variates. The marginals of the Wigner transform are \cite{Athan,Al-Ta} 
\begin{subequations} 
	\begin{eqnarray}
		\int_{k}W(x,k,t)~dk&=&\lvert q(x,t)\rvert^2,\label{e4}\\
		\int_{x}W(x,k,t)~dx&=&\lvert \hat q(k,t)\rvert^2.
	\end{eqnarray}
\end{subequations}
$\hat{q}$ is the Fourier transform of $q$ and the integrations over $k$ and $x$ have the limits $-\infty$ to $+\infty$. Having these marginals Wigner transform can be called as a distribution function. However, it is widely known as  quasi-distribution function as it takes negative values in certain domains of phase space \cite{Vogel}.\\
(ii) We also recall the following property \cite{Athan}
\begin{equation}
	W[q_x,q]=\Big(2\pi ik+\frac{1}{2}\frac{\partial}{\partial x}\Big)~W[q,q].\label{e6}
\end{equation} 	
This property helps us to write  the Wigner transform of a term involving derivatives into derivatives of Wigner transform.

\subsection{\label{ss22} Wigner-SSE}
\par Using the definition (\ref{e2}) and imposing the  properties (i) and (ii) in  both the SSE and its complex conjugate equation, we can represent the  evolution of Eq. (\ref{e1}) in phase space in the form
\begin{eqnarray}
	W_t[q,q]&=&\frac{i}{2}\big( W[q_{xx},q]-W[q,q_{xx}]\big)+~i\big( W[\lvert q\rvert^2q,q]-W[q,\lvert q\rvert^2q]\big)+~ \gamma\big(W[q_{xxx},q]+W[q,q_{xxx}]\big)\nonumber\\
	&&+~3\gamma \big(W[(\lvert q\rvert^2)_xq,q]+W[q,(\lvert q\rvert^2)_xq]\big)+~6\gamma \big(W[\lvert q\rvert^2q_x,q]+W[q,\lvert q\rvert^2q_x]\big).
	\label{e7}
\end{eqnarray}
We note that a similar type of construction has also been made for the NLS equation, Hirota equation and Alber equation, see for example, Refs. \cite{Athan,Al-Ta,Athan2}. Substituting Eqs. (\ref{e4}) and (\ref{e6}) into Eq. (\ref{e7})  we can formulate the Wigner-SSE in the form (see  Appendix \ref{AppA} for details)
 \begin{eqnarray}
		W_t&=&-12{\pi}^2k^2\gamma~W_x~+~3\gamma\int_{\Lambda,\eta,s}e^{-2\pi i\eta s}\Big[W\big(x+\frac{s}{2},\Lambda,t\big)+W(x-\frac{s}{2},\Lambda,t)\Big]d\Lambda~ds~W_x(x,k-\eta,t)~d\eta\nonumber\\
		&&-4\pi k~W_x+i\int_{\Lambda,\eta,s}e^{-2\pi i\eta s}\Big[W\big(x+\frac{s}{2},\Lambda,t\big)-W(x-\frac{s}{2},\Lambda,t)\Big]d\Lambda~ ds~W(x,k-\eta,t)~d\eta\nonumber\\
		&&+12\pi i\gamma\int_{\Lambda,\eta,s}e^{-2\pi i\eta s}\Big[W\big(x+\frac{s}{2},\Lambda,t\big)-W(x-\frac{s}{2},\Lambda,t)\Big]d\Lambda~ ds~(k-\eta)~W(x,k-\eta,t)~d\eta\nonumber\\
		&&+~\frac{\gamma}{4}~W_{xxx}~+~3\gamma\int_{\Lambda,\eta,s}e^{-2\pi i\eta s}\bigg[\frac{\partial W\big(x+\frac{s}{2},\Lambda,t\big)}{\partial\big(x+\frac{s}{2}\big)}+\frac{\partial W\big(x-\frac{s}{2},\Lambda,t\big)}{\partial\big(x-\frac{s}{2}\big)}\bigg]~d\Lambda~ds~W(x,k-\eta,t)~d\eta.
		\label{e8}
	\end{eqnarray}

Equation (\ref{e8}) is a nonlinear PDE in the Wigner transform $W(x,k,t)$. It is not possible to solve the nonlinear PDE (\ref{e8}) analytically. As our aim is to analyze the stability of the system, we move on to analyze the stability of the system (\ref{e8}) against the applied perturbation. This investigation will provide the necessary information about the considered system is stable or not. 

\section{\label{S3}Linear Stability Analysis}
\par In the case of MI, the initial plane wave is modulated by applying a weak perturbation. As time goes on, the perturbation applied on the background starts to grow exponentially. This defines the instability of the system. Sometimes, the perturbation may not exhibit exponential growth marking the stability of that system. In our study we intend to investigate the instabilities that arise from  spectral background.  To achieve this task, we choose  Penrose stability criterion, introduced by Penrose in Ref. \cite{Pen60}. 
\par  Penrose analyzed the stability of a plasma whose dynamics is governed by Vlasov-Poisson equation and framed a criterion as follows : ``If the solution of Vlasov-Poisson equation grows exponentially with time as $t\rightarrow\infty$, as far as linearization is justified, then that solution is unstable which means that the plasma is physically unstable" \cite{Pen60}.  
\par Let us assume that the Wigner-SSE solution describes a homogeneous spectrum. Now let us feebly perturb the solution spectrum in the form 
\begin{equation}
	W(x,k,t)=F(k)~+~\Gamma~ w(x,k,t),
	\label{e9} 
\end{equation}
where $\Gamma\ll1$ whose post factor  denotes the perturbation and $F(k)$ is the Fourier spectrum which represents the homogeneous background. Substituting the assumed solution (\ref{e9}) into (\ref{e8}), we arrive at the following equation, namely the linearized Wigner-SSE, in the order $\Gamma$, that is 

	\begin{eqnarray}
		w_t&=&-~12\pi^2k^2\gamma~w_x ~+~3\gamma\int_{\eta,s}e^{-2\pi i\eta s}~2A^2~ds~w_x(x,k-\eta,t)~d\eta
		\nonumber\\
		&&-4\pi k~w_x+~i\int_{\Lambda,\eta,s}e^{-2\pi i\eta s}\Big[w\big(x+\frac{s}{2},\Lambda,t\big)-w\big(x-\frac{s}{2},\Lambda,t\big)\Big]d\Lambda~ds~F(k-\eta)~d\eta\nonumber\\
		&&+~12\pi i\gamma\int_{\Lambda,\eta,s}e^{-2\pi i\eta s}\Big[w\big(x+\frac{s}{2},\Lambda,t\big)-w\big(x-\frac{s}{2},\Lambda,t\big)\Big]d\Lambda~ds~(k-\eta)~F(k-\eta)~d\eta\nonumber\\
		&&+~\frac{\gamma}{4}~w_{xxx}~+~3\gamma\int_{\Lambda,\eta,s}e^{-2\pi i\eta s}\bigg[\frac{\partial w\big(x+\frac{s}{2},\Lambda,t\big)}{\partial\big(x+\frac{s}{2}\big)}+\frac{\partial w\big(x-\frac{s}{w},\Lambda,t\big)}{\partial\big(x-\frac{s}{2}\big)}\bigg]~d\Lambda ds~F(k-\eta)~d\eta.
		\label{e10}
	\end{eqnarray}
\par The solution of the linearized Wigner-SSE (\ref{e10})  may  either grow or decay in time as $t\rightarrow\infty$. We need an arbitrarily growing solution which is expected to burstforth as a RW. Hence we consider a solution of Eq. (\ref{e10})  in the form
\begin{equation}
	w^{(\xi)}=C^{(\xi)}(k)e^{i(\xi x-\Omega^{(\xi)}t)},
	\label{e11}
\end{equation} 
with $\xi$ is the  modulating wave number which has to be considered as a real parameter, $\Omega^{(\xi)}$ is the modulating frequency which is a complex quantity and $C^{(\xi)}(k)$ is the amplitude of the modulated wave.  We choose this form of solution due of the following reasons: (i) the solution (\ref{e11}) is valid even for an unbounded system ($x\rightarrow\pm\infty$) with the choice $\xi$ is real and (ii) when $t=0$, Eq. (\ref{e11}) neither grows nor decays but when $t>0$, one may observe that the solution either grows or decays depending on the sign of $\text{Im}(\Omega^{(\xi)})$. The linearization withstands as long as the perturbation becomes comparable with the background. By identifying the positive $\text{Im}(\Omega^{(\xi)})$ we can  determine the  unstable solution.  In the following subsection, we derive an expression for the occurrence of  unstable solutions.
\subsection{\label{SS3A}Analytical criterion}
From (\ref{e11}) it is clear that the solution becomes unstable when $\text{Im}(\Omega^{(\xi)})$ is positive. Based on this observation we formulate a condition for the occurrence of  unstable modes.  We recall here that in his studies  Penrose had proposed an integral  dispersion relation as an analytical criterion for the plasma instability \cite{Pen60}. We also formulate a criterion involving an integral using the marginal property of Wigner transform. 
\par Substituting Eq. (\ref{e11}) in the linearized Wigner-SSE (\ref{e10}), the spatial derivatives that appear in the first and fourth term on the right hand side of Eq. (\ref{e10}) can be evaluated straightforwardly. The function $e^{i(\xi x-\Omega^{(\xi)}t)}$ which appears inside the integrals on the right hand side is independent of the variables of integration, so we can take it outside. Now  one can remove this function from the entire equation since it appears in all the terms. On combining the remaining terms suitably and identifying the integrals with delta function, we can rewrite Eq. (\ref{e10})  in an integral form involving delta function in the variable $s$ which is centred at $\frac{\xi}{4\pi}$. Now integrating over $s$, Eq. (\ref{e11}) can  be brought to the form
{\small	\begin{eqnarray}
		\frac{C^{(\xi)}(k)}{\int_{\Lambda}C^{(\xi)}(\Lambda)d\Lambda}&=&\big[-i\Omega^{(\xi)}+4\pi ik\xi+\frac{\gamma}{4}i\xi^3+12i\pi^2k^2\xi\gamma-6iA^2\xi\gamma \big]^{-1}\nonumber\\
		&&\times\quad\int_{\eta}\Bigg[\big(i+3i\xi\gamma+12\pi i\gamma(k-\eta)\big)\delta\Big(\eta-\frac{\xi}{4\pi}\Big)-\big(i-3i\xi\gamma+12\pi i\gamma(k-\eta)\big)\delta\Big(\eta+\frac{\xi}{4\pi}\Big)\Bigg]F(k-\eta)d\eta.
		\label{B1}
\end{eqnarray}}
\par Implementing delta identities we can evaluate the integral over $\eta$ that  appear on the right hand side of (\ref{B1}) from which we can obtain an expression for $C^{(\xi)}(k)$. However, after performing the $\eta$ integration, we once again integrate the resultant equation with respect to $k$ on both sides in order to  obtain an analytical condition. This action yields
{\small
	\begin{eqnarray}
		\int_{k}\frac{C^{(\xi)}(k)}{\int_{\Lambda}C^{(\xi)}(\Lambda)d\Lambda}dk&=&\int_{k}\Big[-\Omega^{(\xi)}+4\pi k\xi+\frac{\gamma\xi^3}{4}-6\gamma\xi A^2+12\pi^2k^2\xi\gamma\big]^{-1}\nonumber\\
		&&\times\Bigg[\Big[1+3\xi\gamma+12\pi\gamma\Big(k-\frac{\xi}{4\pi}\Big)\Big]F\Big(k-\frac{\xi}{4\pi}\Big)-\Big[1-3\xi\gamma+12\pi\gamma\Big(k+\frac{\xi}{4\pi}\Big)\Big]F\Big(k+\frac{\xi}{4\pi}\Big)\Bigg]~dk.
		\label{B2}
\end{eqnarray}}
\par In Eq. (\ref{B2}), as far as the left hand side is concerned,  the numerator can be obtained just by differentiating  the denominator. Hence evaluating the integral on the left hand side, we obtain a  logarithmic function whose argument is the denominator of Eq. (\ref{B2}). Upon evaluating  the integral in the denominator (argument of the logarithmic function) we obtain a constant. Since logarithmic value of a constant is also a constant, we choose the value of that constant as one just for simplicity. As a result we arrive at the following expression
{ \begin{eqnarray}
		1&=&\int_k\frac{\Big[1+3\gamma\xi+~12\pi\gamma~(k-\frac{\xi}{4\pi})\Big]F\big(k-\frac{\xi}{4\pi}\big) - \Big[1-3\gamma\xi+12\pi\gamma~\big(k+\frac{\xi}{4\pi}\big)\Big]F\big(k+\frac{\xi}{4\pi}\big)}{-\Omega^{(\xi)}+4\pi\xi k+\frac{\gamma\xi^3}{4}-6\gamma\xi A^2+12\gamma\pi^2k^2\xi}~dk.
		\label{e12}
\end{eqnarray}}
\par Expression (\ref{e12}) is nothing but the Penrose instability condition. Solutions of Eq. (\ref{e12})   constitute the solutions of linearized Wigner-SSE (\ref{e10})\cite{Athan}. The  solutions of  Eq. (\ref{e12}) are composed of  real  and imaginary part (since $\Omega^{(\xi)}$ is a complex parameter). In the solution we can ignore the imaginary part as it  cannot be  interpreted physically. Hence we concentrate only on the real part and identify its role in making the instability.

\subsection{\label{SS3B}Instability arising out of narrow spectral background} 
\par Upon analyzing Eq. (\ref{e12}), we can perceive the complexity in computing the integrals appearing on the right hand side. The Penrose criterion (\ref{e12}) can be numerically computed for any type of background. We recall here  that the realistic spectra, when become narrow, switch to instability from stable state and vice versa happens when they become broader\cite{Alber}. Hence in order to achieve our task, we consider only the narrow spectrum as background . 
\par Let the background be a delta Fourier spectrum centred at $k_0=\frac{\xi}{4\pi}$, that is $F(k\pm\frac{\xi}{4\pi})=A^2 \delta\big(k\pm\frac{\xi}{4\pi}\big)$. With this choice Eq. (\ref{e12}) becomes 
{\begin{eqnarray}
		1&=&\int_k\frac{A^2\Big[1+3\gamma\xi+12\pi\gamma(k-\frac{\xi}{4\pi})\Big]~\delta\big(k-\frac{\xi}{4\pi}\big)  -  A^2\Big[1-3\gamma\xi+12\pi\gamma\big(k+\frac{\xi}{4\pi}\big)\Big]\delta\big(k+\frac{\xi}{4\pi}\big)} {-\Omega^{(\xi)}+~4\pi\xi k+\frac{\gamma\xi^3}{4}-~6\gamma\xi A^2+~12\gamma\pi^2k^2\xi}~dk.
		\label{e13}
\end{eqnarray}}
\par By executing Dirac identities, we can evaluate both the integrals in Eq. (\ref{e13}). Upon rearranging the resulting expression, we find  
{\begin{equation}
		{\Omega^{(\xi)}}^2+(18\gamma A^2\xi-2\gamma\xi^3)\Omega^{(\xi)}=\frac{\xi^4}{4}-\xi^2A^2-\gamma^2(\xi^6+72\xi^2A^4-18A^2\xi^4).\label{Omr}
\end{equation} }
Equation (\ref{Omr}) is a quadratic equation in $\Omega^{(\xi)}$ whose  roots are given by
{\begin{equation}
		\Omega^{(\xi)}=\frac{\xi}{2}\Big[2\gamma\xi^2-18\gamma A^2\pm i\sqrt{4A^2-\xi^2-36\gamma^2 A^4}\Big].
		\label{e14}
\end{equation}}
\par From Eq. (\ref{e14}), it is clear that  when the  imaginary part of the complex temporal frequency is positive then the solution of the linearized Wigner-SSE  exhibit exponential growth in time (see Eq. (\ref{e11})). The existence of $\text{Im}(\Omega^{(\xi)})$ stands on the fact that the quantity $4A^2-\xi^2-36\gamma^2A^4$ should be greater than zero which in turn provides the maximum and minimum values of $\xi$ in the form
\begin{equation}
	\xi_{\text{min}}:=-2A\sqrt{1-9\gamma^2A^2};~~ \xi_{\text{max}}:=2A\sqrt{1-9\gamma^2A^2}\label{e15}
\end{equation} 
The choice $\xi=0$ fixes the solution as neither a growing one nor a decaying type. Substituting  Eq. (\ref{e14}) in Eq. (\ref{e11}) we can isolate the unstable solution of Wigner-SSE in the form
\begin{equation} 
	w^{(\xi)}=C^{(\xi)}(k)e^{i\Big(\xi x-\frac{\xi}{2}\big[2\gamma\xi^2-18\gamma A^2\pm i \sqrt{4A^2-\xi^2-36\gamma^2A^4}\big]t\Big)}.
	\label{e16}
\end{equation}
\par The  unstable solution  (\ref{e16}) of the  linearized Wigner-SSE (\ref{e10})  exhibits  periodicity in space. One may notice that the solution (\ref{e16}) is given only in partial form and  the  complete expression  can be seen only when we substitute the expression $C^{(\xi)}(k)$ into it. 
The instability of the system shows that it can admit solutions like breathers and RWs. Among these two types of structures, we prefer to investigate the RWs because we have an appropriate marginal to construct them. In the following section, we  identify the region in which RW can originate.
\section{\label{S4}Localized Instabilities}
\par In the study of RWs, most of the efforts have been made  to construct RW solutions  of different orders for the considered system.  To understand the RW occurrences and their properties, attempts  have been made to model them in the form of rational solution for the concerned nonlinear partial differential equation. As far as  SSE is concerned, to begin, first order RW solution was reported by analyzing its  spectral problem, see Ref. \cite{UbanLet}. Later, it has been shown that RW  is one of the limiting cases of  soliton solution on a plane wave background \cite{UbanRev}.  Subsequently  higher order RW solutions have  been constructed for the SSE \cite{Gmu,Mu}. Differing from the earlier works, in this paper, we identify a  region in which RW can occur in the  SSE and calculate the  amount of its localization. 

\subsection{\label{SS41}Localization in space}
\par From the unstable solution (\ref{e16}) we can identify a localized structure. To create a localized wave packet, we superpose the waves in such a way that within the desired region the waves of marginally varying wavelengths in amplitudes and in phases  enhance their overlapping in a constructive fashion  and they destruct each other  elsewhere.
\par Since  Eq. (\ref{e10}) is a linear one, we can superpose  various unstable solutions $(w^{(\xi)})$ of it continuously. 
 The superposed solution can then be obtained through an integration, that is 
\begin{eqnarray}
	w(x,k,t)&=&\text{Re}\int^{(\xi)}e^{i\Big(\xi x-\frac{\xi}{2}\big[2\gamma\xi^2-18\gamma A^2\pm i\sqrt{4A^2-\xi^2-36\gamma^2A^4}\big]t\Big)}~ C^{(\xi)}(k)~B(\xi)~d\xi,
	\label{e18}
\end{eqnarray}
where $B(\xi)$ is the weight of each unstable solution which can have an arbitrary form \cite{Athan}.
\par  Though  we have brought localization to the instabilities yet we do not know which of these localized instabilities can be considered as RW. This situation can be tackled with the help of  marginal property of the Wigner transform, Eq. (\ref{e4}),  which can measure the distribution of the  wavefunction in both space and time. From Eq. (\ref{e4}) and $W(x,k,t)\approx F(k)+\Gamma w(x,k,t)$, we have
\begin{eqnarray}
	\lvert q(x,t)\rvert^2=\int_{k} W(x,k,t) dk \approx\int_{k} [F(k)+\Gamma w(x,k,t)] dk.
	\label{e19}
\end{eqnarray}
Evaluating the integral, we find
\begin{equation}
	\lvert q(x,t)\rvert\approx\sqrt{A^2+\Gamma L(x,t)}.
	\label{e20}
\end{equation}
\par The first term inside the square root, that is $A^2$, represents the area under the spectral curve $F(k)$ (comes out from the integral  involving $F(k)$), has nothing  to do with space and time. The localization can be determined from the second term, $L(x,t)$, where
\begin{equation}
	L(x,t)=\int_{k}w(x,k,t)~dk.
	\label{e21}
\end{equation}
\par  Upon evaluating  (\ref{e21}) we can locate the region where the constructive superposition occurs. Substituting back the superposed unstable solution in Eq. (\ref{e21}), we find
\begin{eqnarray}
	L(x,t)&=&\text{Re}\int_{k,\xi}e^{i\Big(\xi x-\frac{\xi}{2}\big[2\gamma\xi^2-18\gamma A^2\pm i\sqrt{4A^2-\xi^2-36\gamma^2A^4}\big]t\Big)}\times~B(\xi)~C^{(\xi)}(k)~d\xi~dk.
	\label{C1}
\end{eqnarray}
\par  Substituting the expression  $C^{(\xi)}(k)$ (which is found through Eq. (\ref{B1})), in  the preceding equation, we come across a double integral in which the integration of $k$ is nothing but the  Penrose instability condition  (\ref{e12}). Replacing the integral involving $k$ by Eq. (\ref{e12}), we end up at
{\begin{eqnarray}
		L(x,t)&=&\text{Re}\int_{\xi}e^{i\Big(\xi x-\frac{\xi}{2}\big[2\gamma\xi^2-18\gamma A^2\pm i\sqrt{4A^2-\xi^2-36\gamma^2A^4}\big]t\Big)}\times ~B(\xi)~d\xi.
		\label{C2}
\end{eqnarray}  }
\par  We evaluate the integral  (\ref{C2}) to get the localization in $x$ since the superposition of periodic solutions leads to spatial overlapping of waves. We separate the spatial part from the  temporal part with the help of Fourier convolution theorem. Let 
\begin{equation}
	\tilde B_{t}(\xi)=B(\xi)e^{i\frac{\xi}{2}\big[2\gamma\xi^2-18\gamma A^2\pm i\sqrt{4A^2-\xi^2-36\gamma^2A^4}\big]t}.
	\label{C3}
\end{equation}
\par To proceed further  we  introduce a function $(X)$ which acts as a  window function (like the one used  in signal processing) \cite{KMM}. Upon multiplying the window function by another function we get one within the given interval and  zero outside the interval. We can define such a function within the allowed interval of $\xi$ so that the function $X$ makes the outcome zero whenever $\xi$ fails to satisfy the relation (\ref{e15}). Through this way Eq. (\ref{C3}) can be brought to the form
{\begin{eqnarray}
		L(x,t)&=&\text{Re}\int_{\lvert\xi\rvert<2A\sqrt{1-9\gamma^2A^2}}e^{i\xi x}\tilde B_{t}(\xi)~ X_{\big[-2A\sqrt{1-9\gamma^2A^2},2A\sqrt{1-9\gamma^2A^2}\big]}~d\xi.
		\label{C4}
\end{eqnarray}}
\par We multiply and divide the right hand side  by a constant $2\pi$ just before the term  $\xi$ so that Eq. (\ref{C4}) becomes
{\begin{eqnarray}
		L(x,t)&=&\text{Re}2\pi\int_{\lvert\xi\rvert<\mathbb{R}}\tilde B_{t}\Big(2\pi\frac{\xi}{2\pi}\Big)e^{2\pi i\frac{\xi}{2\pi} x}~X_{\big[-\frac{A}{\pi}\sqrt{1-9\gamma^2A^2},\frac{A}{\pi}\sqrt{1-9\gamma^2A^2}\big]}~\frac{d\xi}{2\pi}.\label{C5}
\end{eqnarray}}
\par The right hand side of Eq. (\ref{C5}) can be  interpreted as an inverse Fourier transform of the product of two functions, namely $\tilde B_t$ and $X$, that is
{\begin{eqnarray}
		L(x,t)&=&\text{Re}2\pi~\mathcal{F}_{\xi\rightarrow x}^{-1}\bigg[\tilde B_{t}\Big(2\pi\frac{\xi}{2\pi}\Big)~X_{\big[-\frac{A}{\pi}\sqrt{1-9\gamma^2A^2},\frac{A}{\pi}\sqrt{1-9\gamma^2A^2}\big]}\bigg].
		\label{C6}
\end{eqnarray}}
\par Let $\mathcal{F}^{-1}$ represent the inverse Fourier transform. With the help of Fourier convolution theorem  we can rewrite  Eq. (\ref{C6}) in the form
{\begin{eqnarray}
		L(x,t)&=&\text{Re}2\pi~\mathcal{F}_{\xi\rightarrow x}^{-1}\Big[\tilde B_{t}\Big(\frac{\xi}{2\pi}\Big)\Big]*_{x}~\mathcal{F}_{\xi\rightarrow x}^{-1}\Big[X_{\big[-\frac{A}{\pi}\sqrt{1-9\gamma^2A^2},\frac{A}{\pi}\sqrt{1-9\gamma^2A^2}\big]}\Big],
		\label{C7}
\end{eqnarray}} 
where $*_{x}$ denotes convolution in $x$. The term which immediately follows the convolution symbol in Eq. (\ref{C7}) acts like the Inverse Fourier transform of a rectangular window function $X$. Hence, upon integration, it yields
{\begin{eqnarray}
		L(x,t)&=&4A\sqrt{1-9\gamma^2A^2}~\mathcal{F}_{\xi\rightarrow x}^{-1}\Big[\tilde B_{t}\Big(\frac{\xi}{2\pi}\Big)\Big]*_{x}~\frac{\sin(2A\sqrt{1-9\gamma^2A^2}~x)}{2A\sqrt{1-9\gamma^2A^2}~x}.
		\label{C8}
\end{eqnarray}}
\par The function $(\sin(2A\sqrt{1-9\gamma^2A^2}x))/(2A\sqrt{1-9\gamma^2A^2}x)~$ refers the spatial localization obtained by  superposition of unstable solutions.  From Eq. (\ref{C8}), the actual value of  $L(x,t)$ can be identified  by taking the convolution in $x$.   After the convolution $L(x,t)$ takes  the form \cite{Athan}
{\begin{equation}
		L(x,t)=\int_{v}b(x-v,t)~\frac{\sin(2A\sqrt{1-9\gamma^2A^2}v)}{2A\sqrt{1-9\gamma^2A^2}v}~dv,
\end{equation}}
where 
\begin{eqnarray}
	b(x)&=&4A\sqrt{1-9\gamma^2A^2}~\mathcal{F}_{\xi\rightarrow x}^{-1}\Big[ \tilde B_t\Big]\Big(\frac{x}{2\pi}\Big).~~~~~~~~~~
\end{eqnarray}
\par In the above calculations, we have not assumed any exact form for $B(\xi)$ since it does not contribute anything while calculating the space localization. As the integration in (\ref{C7}) is performed with respect to the modulating wavenumber, the result describes the space localization. The main lobe of the calculated $sinc$ function refers the maximum spatial localization which depends only on $\xi_{max}$.   No more localization is possible through the superposition of unstable solutions of the Wigner-SSE for the same $\xi_{max}$. Hence, for a particular  $\xi_{max}$ (\ref{e15}), the localization will be $\dfrac{4}{2A\sqrt{1-9\gamma^2A^2}~x}$ \cite{Athan}.

\subsection{\label{SS42}RW criterion}
\par  We note that not all localized instabilities can be considered as RW. As we have mentioned in the introduction, ``the RW can emerge within the region $\lambda_0$" \cite{Athan}. 
\par In other words, for a narrow spectral background, the superposed instabilities of Wigner-SSE, in order to describe a RW, should satisfy the condition
\begin{eqnarray}
	\lambda_{0}&\geq&\frac{4}{2A\sqrt{1-9\gamma^2A^2}}.
	\label{e22}
\end{eqnarray}
\par The inequality (\ref{e22}) relates two physical entities, namely wavelength and region of localization. This inequality is satisfied only when the quantity inside the square root $(1-9\gamma^2A^2)$ is positive. This happens only in the case 
\begin{equation}
	A<\pm\frac{1}{3\gamma}.\label{e23}
\end{equation}
\begin{figure*}
	\includegraphics[width=1\linewidth]{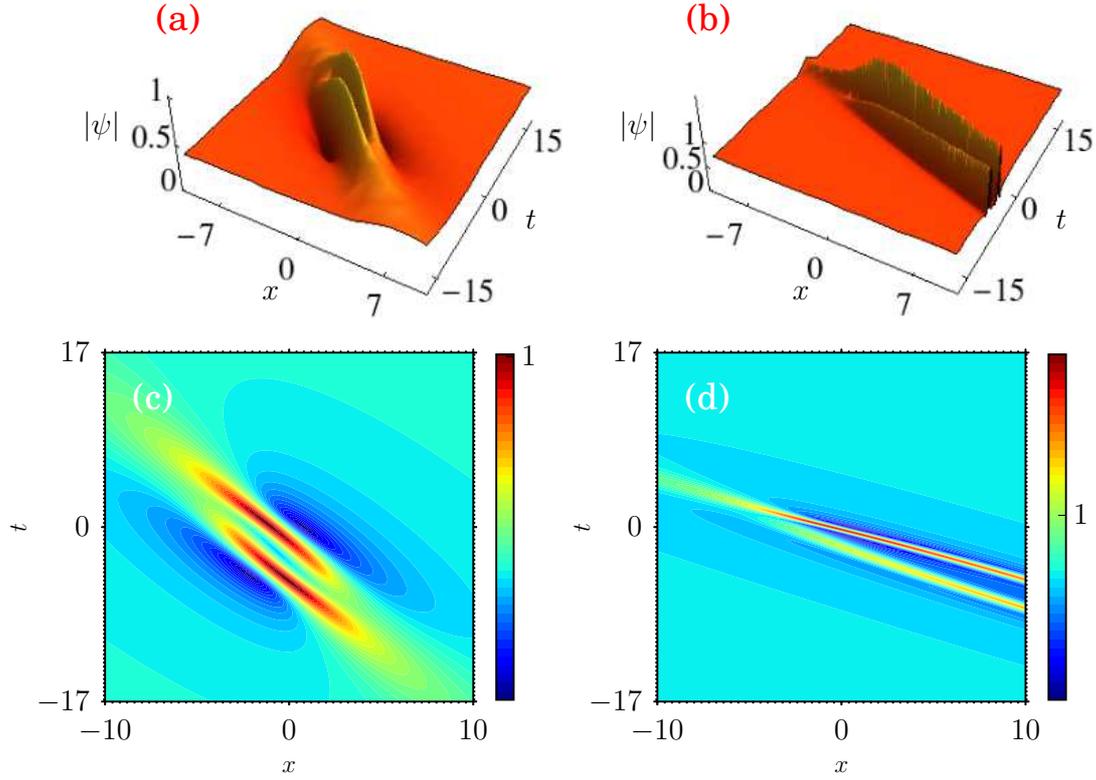}
	\caption{(a) First order RW solution of Sasa-Satsuma equation for $A=0.38$, $\gamma=0.45$. (b) The same for $A=0.76$, $\gamma=0.45$. (c) and (d) are contour plots of (a) and (b) respectively.}
	\label{f1}		
\end{figure*}
\par  So whenever the condition (\ref{e23}) is satisfied one can expect the formation of RW from the considered narrow background. To demonstrate our result graphically, we recall the exact first order RW solution of Sasa-Satsuma equation reported in Ref. \cite{UbanLet}, that is  
{\begin{equation}
		q (x,t)=-\frac{\lambda}{2\gamma}\bigg(1-\frac{\tau-\tau^*}{\lambda}P\bigg)~\exp\Bigg[i\bigg(\frac{k}{2\gamma}x+\frac{\omega}{8\gamma^2}t\bigg)\Bigg]\label{e24}
\end{equation}}
where 
\begin{equation}
	\omega=2\lambda^2-k^2+(6k\lambda^2-k^3)
\end{equation}
and 
{\begin{equation}
		P=\frac{\lvert v\rvert \text{Re}[\tau](\tau v^* \alpha +\tau^* v \beta^*)+(\tau \lvert \alpha\rvert^2+\tau^*\lvert \beta\rvert^2)(\tau^*v^*\alpha+\tau v \beta^*)}{\lvert\tau\rvert^2(\lvert v\rvert^2+\lvert \alpha\rvert^2+\lvert \beta\rvert^2)-\lvert v^2+2 \beta\alpha\rvert^2 \text{Im}[\tau]^2},
\end{equation}}
in which
\begin{subequations}
	{\begin{eqnarray}
			\tau&=&\pm \frac{i\sqrt{9\lambda^2(9\lambda^2+10Q^2)+3\lambda(9\lambda^2-4Q^2)^{\frac{3}{2}}-2k^4}}{3\sqrt{2}Q} \quad \quad~\\
			v&=&\Big(\frac{u_{21}}{2}t-2\gamma x\Big);\quad \beta=\Big(\frac{v}{N_1}+\frac{12\gamma^2}{N_1^2}\Big);\quad \alpha~=~3\lambda\Big(\frac{v}{N_2}-i\frac{12\gamma^2}{N_2^2}\Big); \\ \nonumber\\   
			N_1&=&Q+r-\tau; \quad N_2~=~Q-r+\tau;\\
			r&=&\bigg(\frac{n}{2}+\frac{2(Q^2+18\lambda^2+3\tau^2)}{3 n}\bigg);\quad
			n=(-i+\sqrt{3})\Big([(Q^2-9\lambda^2-\tau^2)\tau]^{\frac{1}{3}}\Big);
	\end{eqnarray}}
	{		\begin{eqnarray} 			
			u_{21}&=&\frac{9(m-6\lambda^2)\tau^4+3 m(m-1-18\lambda^2)\tau^2+m^3}{3(2\tau^2+r\tau+m^2)}; \\ \nonumber\\
			m&=&\frac{(Q^2-3-36\lambda^2)}{3};  ~~~  Q=1+3k; \quad \lambda=2\gamma A .\label{e27h}
	\end{eqnarray} }
\end{subequations}
\par We plot the solution (\ref{e24}) in Fig. \ref{f1}.  When the solution obeys the condition (\ref{e23}) we obtain a double peaked RW  as shown in Figs. \ref{f1} (a) and (c).  If the condition (\ref{e23}) is violated the localization of RW is distorted which is demonstrated  in Figs. \ref{f1} (b) and (d). The plots are drawn for  $k=0$.  
\par The obtained result  (\ref{e23}) matches with the one reported in Ref. \cite{UbanLet}, in which the authors have pointed out that RW can exist for $k=0$ with a constraint that the amplitude parameter obeys a particular relation (see Eq. (12) in Ref. \cite{UbanLet}). Because of this the wavenumber disappears from the amplitude parameter relation. The result which we have obtained from the narrow spectral background  corresponds to the case $k=0$ reported in  Ref. \cite{UbanLet}. In our case, marginal property (vide Eqs. (\ref{e4}) and (\ref{e21})) enforces the wavenumber to disappear from the condition (\ref{e23}). 

\section{\label{S5}Conclusion}
\par In this paper, we have calculated the region in which RW can emerge in the SSE and the exact value of localization. To achieve this task, we have rewritten the SSE in phase space using Wigner transform and its properties. We have performed Penrose stability analysis for the SSE written in phase space. The Wigner-SSE came out as a nonlinear PDE. We have carried the stability analysis for the Wigner-SSE. It is  known that an arbitrary growth in the solution is a necessary  but not a sufficient condition for a RW to occur. By utilizing  the linear superposition principle we have made the instabilities localized. With proper justification given for the linearized form of solution, we compared these localized instabilities with the definition of RW from which we deduced  a sufficient condition for the existence of RW. We have also demonstrated our result graphically.
\par We note here that our outcome is consistent with the results reported in Refs.  \cite{UbanRev,UbanLet}. In the latter two references the authors have studied RW that arises from the plane wave background whereas in our studies we have focussed our attention on RWs that arise from the narrow background spectrum. Our result shows that RW can exist whenever the system parameter $(\gamma)$ and the amplitude $(A)$ obey a particular amplitude - parameter relation given in Eq. (\ref{e23}). When the system parameter and the amplitude fails to satisfy this relation, the shape of RW gets distorted and eventually the wave disappears. 
\par The analysis which we have carried out in this paper has also been performed on the NLS equation earlier. One may consider the SSE as NLS equation with an extended nonlinearity. From this point of view it is worth to compare the outcome of the present work with the one that performed on NLS equation. In the NLS equation, the authors have found the fundamental lengthscales and timescales for the RW that emerges from  non-dissipative and very narrow spectrum \cite{Athan}. In our work, we have established a relation between amplitude and system parameter for which RW can emerge, from a non-dissipative and very narrow spectrum. Unlike the NLS equation, in SSE, we come across a double peaked RW structure. Our result will be useful to the experimentalists working on higher order nonlinear phenomena.

\appendix

\section*{Appendix: \label{AppA}Construction of Wigner-SSE}

\par In this Appendix, we formulate Wigner-SSE for the Eq. (\ref{e1})  by utilizing  Wigner transform and its properties. Using Fourier convolution theorem, we can rewrite the nonlinear terms in the required form. Using Eq. (\ref{e6}) we can rewrite the group velocity dispersion term as
\begin{equation}
	W[q_{xx},q]-W[q,q_{xx}]=4\pi ik ~W_x[q].
	\label{A1}
\end{equation} 
\par  Similarly, the third order dispersive term in (\ref{e7}) (second term) can be rewritten in the form
\begin{equation}
	W[q_{xxx},q]+W[q,q_{xxx}]=-12\pi^2k^2 ~W_x + \frac{1}{4} ~W_{xxx}.\label{A2}
\end{equation}
\par The self phase modulation term $T_3$ (third term in Eq. (\ref{e7})), can also be rewritten in an exact Wigner form with the help of the convolution theorem of Fourier transform (after replacing $\lvert q\rvert^2=U(x,t)$), that is 
{\begin{equation}
		T_3=\int_{\eta,s}e^{-2\pi i\eta s}~[U\big(x+\frac{s}{2},t\big)
		-U\big(x-\frac{s}{2},t\big)]~ds~W(x,k-\eta,t)~d\eta.
		\label{A3}
\end{equation}}
\par In the self-frequency shift term  $T_4$ (fourth term in Eq. (\ref{e7})), we have the spatial derivative of $U(x,t)$. We keep this term as it is and rewrite this term as
{\begin{equation}
		T_4=\int_{\eta,s}e^{-2\pi i\eta s}\Bigg[\frac{\partial U\big(x+\frac{s}{2},t)}{\partial\big(x+\frac{s}{2}\big)}+\frac{\partial U\big(x-\frac{s}{2},t\big)}{\partial\big(x-\frac{s}{2}\big)}\Bigg]ds~W(x,k-\eta,t)d\eta.
		\label{A4}
\end{equation}}
\par To rewrite the last term ($T_5$), we use the properties (\ref{e4}) and (\ref{e6}). After imposing these properties and through the convolution the last term can be brought to the form 
 
		\begin{eqnarray}
			T_5&=&\int_{\eta,s}e^{-2\pi i\eta s}\Big[U\big(x+\frac{s}{2},t)-U\big(x-\frac{s}{2},t\big)\Big]~ds~2\pi i~(k-\eta)~W(x,k-\eta,t)~d\eta\nonumber\\
			&&+\int_{\eta,s}e^{-2\pi i\eta s}\Big[U\big(x+\frac{s}{2},t)+U\big(x-\frac{s}{2},t\big)\Big]~ds\frac{1}{2}~W_x(x,k-\eta,t)~d\eta.
			\label{A5}
		\end{eqnarray}

\par Replacing the right hand side of Eq. (\ref{e7}) with the above expressions (\ref{A1})-(\ref{A5}), we obtain the Wigner-SSE which is given in Eq. (\ref{e8}) in Sec. \ref{S2} with $U(x,t)=\displaystyle\int_\Lambda W(x,\Lambda,t)~d\Lambda$  owing to the marginal of the Wigner transform.

\section*{Acknowledgments}

MP acknowledges the Bharathidasan University, Tiruchirappalli, for providing a University Research Fellowship (URF). NVP wishes to thank IISc, Bangalore, for providing a fellowship in the form of  Research Associateship. The work of MS forms part of a research project sponsored by NBHM, Government of India.

\textit{The Data that supports the findings of this study are available within the article.
}


\end{document}